# The ATLAS EventIndex

## A BigData Catalogue for All ATLAS Experiment Events

Dario Barberis[1] · Igor Alexandrov[2] · Evgeny Alexandrov[2] · Zbigniew Baranowski[3] · Luca Canali[3] · Elizaveta Cherepanova[4] · Gancho Dimitrov[3] · Andrea Favareto[1] · Álvaro Fernández Casaní[5] · Elizabeth J. Gallas[6] · Carlos García Montoro[5] · Santiago González de la Hoz[5] · Julius Hřivnáč[7] · Alexander Iakovlev[2] · Andrei Kazymov[2] · Mikhail Mineev[2] · Fedor Prokoshin[2] · Grigori Rybkin[7] · José Salt[5] · Javier Sánchez[5] · Roman Sorokoletov[8] · Rainer Többicke[3] · Petya Vasileva[3] · Miguel Villaplana Perez[5] · Ruijun Yuan[7]

## Abstract
The ATLAS EventIndex system comprises the catalogue of all events collected, processed or generated by the ATLAS experiment at the CERN LHC accelerator, and all associated software tools to collect, store and query this information. ATLAS records several billion particle interactions every year of operation, processes them for analysis and generates even larger simulated data samples; a global catalogue is needed to keep track of the location of each event record and be able to search and retrieve specific events for in-depth investigations. Each EventIndex record includes summary information on the event itself and the pointers to the files containing the full event. Most components of the EventIndex system are implemented using BigData free and open-source software. This paper describes the architectural choices and their evolution in time, as well as the past, current and foreseen future implementations of all EventIndex components.

**Keywords** ATLAS experiment · Event catalogue · BigData catalogue · Hadoop · HBase

## Introduction

The ATLAS experiment [1] at the CERN LHC accelerator [2] is in operation since 2009; it collected during the first two LHC data-taking periods (Run 1 between 2009 and 2013 and Run 2 between 2015 and 2018) almost 25 billion physics records ("events"). The particles accelerated by the LHC are grouped into many "bunches" that intersect each other every 25 ns, at a rate of 40 MHz. During each bunch crossing, several independent interactions take place at almost the same time (within 0.5 ns); therefore, the signals left in the detector by the particles produced by those interactions are recorded together as one "event". The average number of these "pile-up" interactions varied from a few in LHC Run 1 to 50–60 during LHC Run 2 and is expected to increase further by the end of LHC Run 3 (2022–2025). In addition to the real events, about three times as many simulated events were generated using Monte Carlo methods.

## ATLAS Data Taking and Data Processing

The LHC accelerator operates in cycles; first the protons or ions are injected into the accelerator rings, then they are accelerated and finally their orbits are modified to bring the counter-rotating particle beams to intersect at the centre of each experimental apparatus. When the number of circulating particles has decreased beyond a certain level, they are extracted and directed towards the "beam dumps", where their energy is absorbed. Each such cycle is called a "fill" for the accelerator. A continuous data-taking session of the detector is usually referred to as a "run", typically lasting from a few hours to just over a day. Runs are further divided into luminosity blocks, short time periods, up to one minute, during which the LHC luminosity (interaction rate) and the detector conditions can be assumed to be constant. Luminosity block identifiers, called LBN (luminosity block number), are integers starting from 1 and incremented as the run progresses.

The detector read-out system and the downstream processing chain cannot cope with the 40-MHz bunch crossing rate. An online selection system ("trigger") is needed to select the events of interest on the basis of combinations of signals in the detector that match the expected signatures for interesting physics processes, and extract their full information for offline processing. The trigger system had a three-level configuration for LHC Run 1: Level 1 (L1), implemented in programmable hardware; Level 2 (L2), that made software-based partial reconstruction within the "regions of interest" marked by Level 1, and finally the Event Filter (EF), analysing the full event. For LHC Run 2 and Run 3, L2 and EF were merged into a single system, the High-Level Trigger (HLT), implemented in software running on commercial processors. In order to keep the possibility to select rare physics processes, events satisfying triggers matching common physics processes or used for monitoring were prescaled, i.e. reduced randomly in number in order to save output bandwidth. The trigger configurations and the prescale factors applied to each trigger type were recorded in the trigger database for each run [3].

Trigger decisions were stored in event data files as trigger masks, where each bit corresponds to the specific set of trigger selections (trigger chains). Depending on the trigger configuration, the same bit in the trigger mask (chain counter) may correspond to a different trigger chain. The relation between chain name and counter is uniquely defined in the trigger database table, indexed by the trigger key (SMK). Trigger keys depend on the run number for the data recorded by the detector and on the production settings for Monte Carlo simulation.

The events recorded by the ATLAS detector are transferred in real time to the CERN computer centre and processed within 48 h from the end of a run using the ATLAS "Tier-0" cluster. This procedure consists of calculating the time-dependent detector calibration and alignment parameters, and then using them to compute for each event the physical quantities of interest for the final analyses. The events can be re-processed when improved reconstruction algorithms or detector calibration and alignment constants become available, usually at the end of each major data-taking period; the result is the creation of additional versions of the same events, some of which replace older versions.

The reconstruction processes take events in "RAW" data format, as produced by the detector read-out system, apply detector calibration and alignments, execute particle reconstruction and identification algorithms and output them in "AOD" (Analysis Object Data) format. The events are then distributed and made available to all ATLAS members through the World-wide LHC Computing Grid (WLCG [4]). The events can be further selected for different analysis purposes, and saved in compressed formats ("derived AOD", or "DAOD") with only the events and contents that are useful for a given analysis. These derivation processes can be run very frequently, even monthly, as the analysis codes evolve in time, resulting in many DAOD versions with a relatively short lifetime as they are normally superseded by newer ones.

Simulated interactions go through a similar processing chain. The outputs of event generators are saved on disk in a common format ("EVNT"). Then the detector and read-out electronics simulation (digitization) processes are run; pile-up interactions are also simulated and added to the main interaction record during digitization. Finally, the same reconstruction and derivation processes as for real data are executed, producing events, respectively, in AOD and DAOD formats.

Groups of statistically equivalent events (real data events collected with the same detector conditions or simulated events produced with the same generator, and processed by the same software versions) are stored in files on disk or on tape. Each file typically contains between 1000 and 10,000 events, depending on the format and balancing the need to avoid too many small files (<1 GB) that would cause data storage problems and at the same time too large files (>10 GB) that could have lower transfer efficiency. Every unique file, regardless of its format, is assigned a distinct GUID (Globally Unique IDentifier [5]), which is used to catalogue and retrieve it. Currently ATLAS has over 100 million files on disk, containing over 400 billion event records. Files are grouped into *datasets* that can be hierarchically assembled into *containers*. The distributed data management system Rucio [6] is used to keep track of each file, dataset and container, including their properties (metadata) and replica locations, as well as to manage the data movements between different storage sites and the CPU farms where the data are processed and analysed.

### Need for an Event Catalogue

Rucio is extremely efficient at managing 100 million files stored in 120 sites world-wide and using over 200 PB of disk space and similar quantities of tape storage space. Rucio allows analysers to easily access data and simulation samples using the existing organization of files in datasets and containers, but it does not store any information about individual events. Nevertheless there is also a need to be able to retrieve different versions of one or a few events, given the often reduced information contained in the last data reduction stages, either to produce nice event displays for publications, or to check in detail if all reconstruction procedures worked as expected.

The EventIndex was designed for this primary use case (quick and direct event selection and retrieval), but the same system can fulfil several other tasks, such as checking the correctness and completeness of data processing procedures, detecting duplicated events that can occur for temporary faults of the data acquisition or processing procedures, studying trigger correlations and the overlaps between selected data streams.

The EventIndex is the second-generation event catalogue for ATLAS. The first-generation catalogue deployed for LHC Run 1 was called the "TAG DB" [7]. Its content was based on the direct import of event-wise information from ATLAS TAG files into an Oracle [8] database to facilitate queries of all events across entire datasets, which could speed up the analysis workflow. TAG files were thumbnails of the AOD format files produced in the final stage of the central production chain. The TAGs contained, in principle, sufficient information for identification and selection of events of interest to most physics analyses for the purpose of subsequent event skimming, i.e. the pre-selection of events based on any of the TAG quantities before accessing the full AOD file content. There were generally two problems with the TAG database:

1. TAG information included immutable content (such as event identification and trigger decisions) as well as mutable content (basic physics quantities such as the number of loose electron candidates and their kinetic information). The inclusion of the mutable content was problematic because in practice, recalibrations and software improvements were made in post-processing or reprocessing steps, which skipped the production of new TAG files, so the TAG DB mutable content became out-of-date (stale) for accurate event selection based on that content. There was no easy way to refresh mutable content without fresh TAG files.

2. The TAG database structure was, for simplicity, dictated by the TAG files because the skimming functionality was based on being able to produce TAG files from the database. This meant that mutable content could not be easily separated from the immutable content, so all content was put into single database tables with hundreds of columns. To enable fast queries based on any of the content every column was indexed, which was a very heavy implementation on the database side, requiring a lot more storage than was ever actually used due to the first problem. While the mutable content became a dead weight on the system, there were many use cases for the immutable content which were utilized.

An inherent problem with the TAG DB was the dependence of the system on the ATLAS production processing chain to produce TAG files with each iteration of recalibration and software improvements which was rarely fulfilled (except in full reprocessing after years of data taking). This fundamental flaw was completely avoided in the EventIndex catalog described in this paper by using a far superior workflow: it deploys its own data collection jobs (not part of the central production chain) and is able to collect event metadata from files at any stage of event processing (not just AODs). We also decided to avoid the mutable content in the new EventIndex since the Run 2 analysis model has its own mechanisms for event skimming and to expand instead on the many use cases utilizing the immutable content as will be described in later sections. Since the data collection and storage were completely revamped, it was decided to explore the (then) new BigData technologies that were becoming increasingly popular and promising in terms of scalability with respect to data volumes, and start developing the new EventIndex in advance of the start of LHC Run 2 [9, 10].

This paper is organized mainly following the data flow through the EventIndex components. "Requirements and Global Architecture" describes the use cases and the overall system architecture; "Data Production" and "Data Collection" explain the selection of datasets to index, the indexing method and the index data collection components; "Data Storage in Hadoop and HBase" and "Data Storage in Oracle" describe the current data storage methods in Hadoop [11], HBase [12] and Oracle. "System Monitoring" and "Operations and Performance" cover system operations and monitoring. Finally, "System Evolution" outlines the developments towards a higher performance system for LHC Run 3 and beyond.

### Requirements and Global Architecture

We begin by briefly giving an overview of the use cases for the collected EventIndex records. Details of the utilities implemented to satisfy these use cases are described in later sections. These cases originated from experience gained from the LHC Run 1 catalogue and from experts within the collaboration who recognized that the information consolidated in the EventIndex system could facilitate investigations across datasets much more easily than through processing event files.

## Use Cases and Functional Requirements

The main use case, and in fact the *raison d'être* of the EventIndex, is the so-called "event picking". Often in the course of a physics analysis it is necessary to retrieve full information about one or a few events, either to generate event displays for publications (see examples in [13]), or to inspect its properties and verify the correctness of its reconstruction procedure. A person who needs to search, select and/or retrieve one or more events out of the many billions of ATLAS events needs a complete catalogue of all events, in all processing versions, including the pointers to the event locations. This catalogue, similarly to the catalogues of (paper) book libraries, needs to contain enough information ("metadata") about each event to be useful for the search, and at the same time it needs to provide reasonably fast queries, at least for the most common cases.

Several other use cases can be served by a complete event catalogue. A second group of use cases are related to data quality assurance. Production consistency and completeness checks can be run, for example counting the number of events in the input and output datasets for any processing task and making sure that all events have been processed, and there is no data duplication in the output datasets. Running indexing jobs on all produced data provides in addition a check that all output files are stored correctly on disk and are available for further analysis.

Further use cases are related to the calculations of overlaps among trigger chains within a given dataset, and and among derived datasets. With the complete trigger information for each event stored in the catalogue, it is possible to count the number of events that satisfied each trigger chain and also measure the overlaps among trigger chains. Similarly, it is possible to select particular events on the basis of combinations of their properties, such as the trigger or the instantaneous luminosity (interaction rate), and then count their occurrences or retrieve them directly for full analysis.

Not all event processing tasks output the same number of events they had in input. The derivation procedures take all fully reconstructed events as input and output only the selected events that are useful for one or a few particular analyses. As there are almost 100 derivations that run on ATLAS events, it is useful to have the possibility to check the amount of overlaps between derivation streams, in order to reduce if/when possible their number and hence the processing time and the disk space for the output datasets.

Differently from the trigger stream overlap checks that are done within a specific dataset, the derivation overlap check involves a number of different but related datasets; related in the sense that all these derived datasets must have been produced from the same parent one.

## EventIndex Record Contents

The EventIndex stores only "immutable" event parameters, i.e. those that do not depend on the processing version, excluding all physics parameters of the simulated or reconstructed events.

In order to satisfy the use cases, each event record needs to contain three blocks of information:

1. *Event identification* Each instance of a given event can be uniquely identified by the combination of run number, event number, trigger stream, data format and processing version; therefore, this information has to be included in each event record. In addition, the data type (real or simulated data), time stamp, LHC conditions, luminosity block number (only for real data) and (for simulated events only) event weight and simulation process identifier are included as they can be useful to trace possible processing problems and for future reference.
2. *Trigger information* Trigger masks for the L1, L2 (only for LHC Run 1) and HLT triggers, the trigger key (SMK, used to decode the trigger masks) and the prescale key (with information on the trigger prescale settings). The SMK can be used together with the trigger database [3] to decode the trigger records of each event and show which trigger chains led to the event being recorded.
3. *Location information* The GUID of the file that contains this event and the internal pointers within that file, for the file that is currently indexed and also for the upstream files in the processing chain (provenance). The GUID can be passed to Rucio to identify, locate and retrieve the file containing a given event in order to extract it or analyse it directly. The provenance record is useful to reduce the number of datasets that have to be indexed; for example, the pointers to the RAW datasets can be obtained by indexing the corresponding AOD datasets.

## Performance Requirements

The catalogue must sustain a record ingestion rate that is at least as large as the real data production rate (1 kHz during LHC Run 2), plus the simulated data processing rate (about the same when averaged over a year). Given the foreseen increase of trigger rates by the end of LHC Run 3 to over 3 kHz and the corresponding increase in simulation production computing power, the catalogue for LHC Run 3 needs to withstand an ingestion rate of 10 kHz at least, allowing for some contingency in case of operation backlogs. The query rate is very small compared to the ingestion rate, but all data are equally important and all queries are different, so it is necessary to have a flat internal structure and caching does not help much. User queries for small data samples are done through client code with a command-line interface, or through web services, requiring that the response times for simple queries be compatible with human response expectations (below 1 s); queries for large amount of data or computations of global counts, trigger overlaps or derived dataset overlaps can be executed as batch processes but need to return their results within (roughly) an hour, and never fail, in order to be useful.

## System Architecture

The information flow through the EventIndex system is linear, so it was natural to match the system architecture to the data flow [10]. One needs first to extract the relevant metadata from the event data files and store them in a central store, which client programs can query to perform their tasks. The EventIndex system is therefore partitioned into a number of components:

- *Data production*. This component takes care of extracting the metadata from each data file as soon as it is produced at CERN or on any of the ATLAS Grid sites, format them for transfer to the central store and send this information to CERN.
- *Data collection*: This component deals with the data transfer infrastructure, the metadata completeness checks for each dataset, assembling the information produced by all files in a given dataset and formatting it for storage, including decoding the trigger information and presenting it in an optimized format for fast searches.
- *Data storage*. This is the core system. It includes the setup of the EventIndex data storage cluster in Hadoop, the code to import the data and internally index them, and the web service providing the command line and graphical interfaces for the clients. As a subset of the EventIndex data are also replicated to an Oracle database for access performance reasons, this component includes also the support for the Oracle store, the data import code and the graphical interface for the users.
- *Monitoring*. All servers and all processes have to be constantly and automatically monitored. This component collects, stores and displays the relevant information, and sends automatic alerts in case of service interruptions or malfunctioning. Regular functional tests are also submitted in order to monitor the performance for the most common use cases.

Figure 1 shows a schema of this global architecture. Thanks to the partitioning and to the clear interfaces between components, it is possible to implement, evolve and upgrade each component independently of the other ones. The data production and data collection components already went through a couple of upgrades; the data storage component was upgraded to a newer base technology in advance of the start of LHC Run 3 in 2022 (see "System Evolution").

This architecture allows the development of additional services that satisfy more complex needs, such as the Event Picking Server [14] that will automate most of the actions needed for event picking (see "System Evolution").

Only a few components depend on ATLAS data structures, namely the Producer, which has to read ATLAS data files, and the storage schema; other experiments could use the same infrastructure by just replacing those components with theirs.

## Data Production

The data production system includes all tasks that are executed at the sites where the datasets to be indexed reside, in order to collect information from the event files and transfer it to the central EventIndex servers.

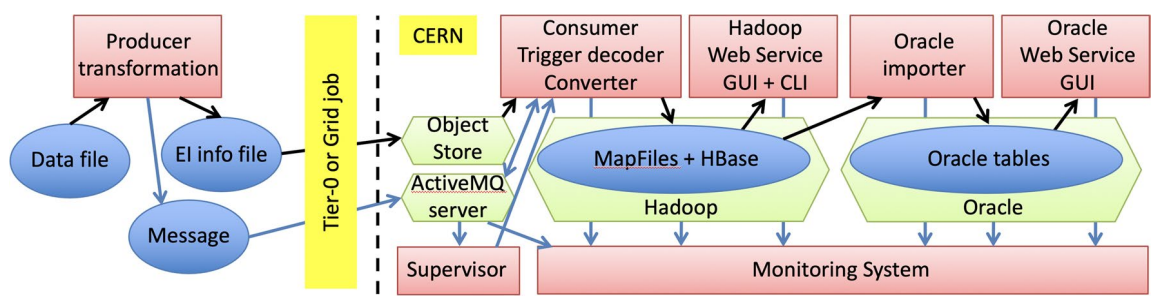

**Fig. 1** Global architecture of the EventIndex system, as implemented at the end of LHC Run 2. The blue ovals indicate temporary or permanent data blocks or files; the green hexagons correspond to different storage technologies. The pink rectangles contain continuously running processes. The black arrows show the flow of EventIndex data; the blue arrows show the flow of information related to data processing. Further details are explained in "System Architecture"

## Dataset Selection

As soon as new ATLAS data are processed on the CERN Tier-0 cluster [15] and the corresponding AOD datasets are available, jobs are launched to extract the EventIndex information for all "physics" datasets; calibration, test and monitoring streams are excluded. From this indexing step, information on the location of each event in RAW format is also extracted.

Many more ATLAS datasets are produced using the ATLAS resources of the WLCG Grid, and other additional resources that can be available from time to time. They include the whole simulation chain, from event generation to detector simulation, then event reconstruction and selection for analysis; real events are also re-reconstructed from time to time on the Grid, and all analysis selections also take place in a distributed fashion.

All AOD datasets are indexed, and for real data all types of derived AODs (DAODs) are indexed too. For the simulated data only some types of DAODs are indexed, if requested by the analysis groups that use them. In addition, all EVNT datasets are also indexed.

Datasets produced on the Grid are selected for indexing according to the following criteria, based on information obtained from the ATLAS metadata database AMI [16]:

- The dataset is marked in AMI as complete and validated for use in analysis or further processing.
- The dataset is marked as long-lived, to avoid indexing transient datasets that are only used in internal steps of the production procedure.
- The dataset is part of a regular production processes and not just used for checks or validations of the software or the trigger configurations.

Selected datasets should then pass additional checks, to exclude datasets that have a "bad" status in Rucio, are known for problems or have other signs of corrupted data that may cause import jobs to crash or result in excessive computing resource consumption.

## Indexing Job Submission

Datasets that were selected for indexing have to be processed by the Production and Distributed Analysis system (PanDA) [17]. PanDA takes the list of new datasets and generates jobs that run a predefined "transformation" (a script containing a sequence of algorithms to be executed on a data file [18]) on the WLCG Grid.

The transformation used and its configuration in general depend on the data format of the dataset and the type of data (simulated or real); for example, the trigger information, which constitutes a large fraction of the EventIndex data, is collected for each event only from datasets in AOD format as it will not change with subsequent processings of the same event. The progress of these jobs can be monitored through a dedicated dashboard; if necessary jobs can be aborted or rerun.

## Producer Transformation

The Producer is in charge of extracting the EventIndex information from the actual input files, store it into temporary files and send them to a central location at CERN. It has to be able to run using the ATLAS production infrastructure on all available production facilities (the Tier-0 cluster at CERN and the WLCG Grid), so it is implemented in a way very similar to standard ATLAS data processing programs using the ATLAS transformation framework running within the Athena software framework [19].

Python was chosen as the implementation language for the Producer code, as it only accesses the header records of each event. The python interfaces to the Athena classes methods written in C++ do not change between releases, so the Producer code can be rather stable.

The producer input can be one or several files in the ATLAS specific ROOT format [20, 21], such as those in AOD, DAOD and EVNT datasets.

The EventIndex transformation class implements all required methods by the Athena framework to initialize a job, execute the event loop (process the events) and finalize the job. The current implementation runs using a serial processing model, so the input data structure (file and event ordering) is preserved without needing further post-processing.

The EventIndex producer has two separate steps, both running within the transformation. In the first step, it reads events (*execute* method), extracts information and saves the relevant information into a temporary file. When all events are read, the second step starts (*finalize* method), in which the output file is transferred to a central store at CERN. Besides the EventIndex information itself, some additional environment and processing information is stored: the PanDA task and job identification, input dataset name, total number of files and events, starting and ending processing times as well as identification (GUID) and number of events for each file read.

This second step provides a good opportunity to check for inconsistencies in ATLAS data files as soon as they are produced. Currently the transformation looks for event uniqueness within each file, so duplicate events that could result from failures in previous processing steps, are detected here allowing quick notification to ATLAS.

## Data Collection

The data collection system receives and validates the information extracted by the producers, assures its completeness, and orchestrates the EventIndex data transfer from the producers that run on the WLCG Grid to the Hadoop cluster at CERN. Depending on the number of processed events, each indexing job produces between 100 kB and a few MB of information to be transferred to the central servers.

### Messaging System

Messaging is a key component of the data collection system. In the original implementation of the producer transformation [22], the output file was serialized and packed into JSON messages, sent to ActiveMQ brokers [23] at CERN using the STOMP protocol [24].

Two different types of messages were used in the messaging based data collection architecture:

1. *Data messages*, containing the produced data. They ranged from 1 to 10 kB and were tagged in a way that all messages from the same producer were consumed by the same consumer. Larger payloads were split into 10-kB chunks and sent as independent messages; the consumer processes were then recombining them into a single file.
2. *Status messages*, allowing the tracking of the indexing processes. They were sent to a different queue, where they were collected. The produced information was validated by means of the status messages.

Although this architecture was reliable and fully functional, there were concerns regarding its ability to cope with peaks of production activities, as all information was kept in the brokers until it was consumed. During peak times, this could lead to a considerable growth in the number of messages that the brokers have to keep until they can be delivered and consumed.

The current system still uses the messaging system for *Control Messages*, which are similar to the original status messages. In this way the amount of data flowing through the ActiveMQ servers was reduced from a few megabytes (dominated by the data messages) to a few tens of kilobytes per job (just the control message).

### Object Store

Alternatives to minimize the impact of the expected increases of data-taking rates on the messaging architecture were investigated [22], resulting in the replacement of data messages by temporary objects written into an Object Store [25]. Figure 2 describes this approach. A temporary object is created by each producer job, containing the information that the producer transformation ("Producer transformation") extracted from the processed files. Once the object is written into the S3 Object Store [26] at CERN, a *Control Message* containing a summary of the information and the URI of the object is sent to a new entity, the EventIndex Data Collection Supervisor ("Supervisor"), that orchestrates all data collection activities.

The information sent through the messaging mechanism has therefore been drastically reduced from several megabytes to tens of bytes for each job, keeping the brokers in a well-performing status. If a producer, for any reason, is not able to write the data into the object store, there is a fallback solution based on the CERN large-data store EOS [27] also at CERN, using the xrdcp protocol [28].

### Index Record Format

Two different file formats were used to store the output from the producer, adapted to the specific needs of the processing.

Initially a SQLite3 [29] format was reused from other Athena tools, with data stored in key:value pairs using only one table with two columns, "key" (TEXT) and "value" (BLOB). "Value" is the serialized representation of a python object using cpickle, so arbitrary objects can be saved and retrieved into the database, allowing a flexible "blackboard" style storage of key:value pairs. The producer transformation used several key: value pairs to store general information like the number of files and events processed, date and time of processing, job and task identification, file GUIDs, input collection name (dataset name), etc. Events were saved consecutively as an ordered tuple with key "Entry N" where N is the entry number. This file format was successfully used by the first producer implementation, but when it was decided that the file was going to be sent "as is" to the S3 Object Store it was soon realized that it was not the best format for the new needs.

The current producer uses a format based on the Google Protocol Buffer [30] with gzip [31] compression. This format allows the consumers to read the files easily and the size reduction achieved by the compression allows faster transfer times and requires fewer resources in the S3 Object Store.

This format contains a Stream of Protocol Buffer (ProtoBuf) messages (SPB) compressed using the gzip library on the fly. The uncompressed file starts with a "magic" fixed_uint32 value (0x6e56c8c7) so it can be identified quickly. Since ProtoBuf messages do not have type information, all messages have extra prepended information to identify the message type; two fixed_uint32 integers containing the type and message version and its length are added before the message itself.

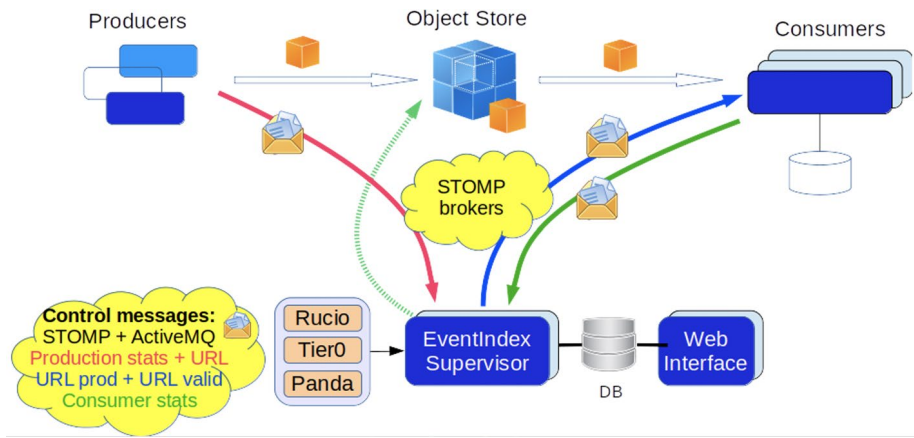

**Fig. 2** Architecture of the EventIndex Data Collection system based on Object Store. The data flow is described in "Supervisor". The thick arrows indicate the flow of EventIndex data from the Producers to the Consumers, going through the Object Store; the red arrow marks the messages sent by the Producers to the Supervisor through the message broker; the light dotted green and the blue arrow correspond, respectively, to the information stored by the Supervisor in the Object Store about the location of all objects related to a given dataset, and the signal sent to the Consumers that all data for a dataset is available in the Object Store; the dark green arrow marks the messages sent back by the Consumers to the Supervisor to signal the completion of a given dataset transfer

The file can contain six different message types: Header, Trailer, BeginGUID, EndGUID, TriggerMenu and EIEvent:

1. Header: contains global information about the processing step, like task and job identifications, input dataset name and start processing time.
2. Trailer: contains global information collected during the processing, like number of files read, total number of events and end processing time.
3. BeginGUID: marks the start of a new input file being processed. These messages contains the input file unique global identifier (GUID), and the start processing time besides some other ATLAS metadata information like the processing version, the stream and project names.
4. EndGUID: marks the end of the input file processing. It contains the number of events read for this file and the end processing time.
5. TriggerMenu: contains the trigger menu used during data taking for the next collection of events. This message is sent once per file read and whenever the trigger menu changes.
6. EIEvent: this is the main part of the EventIndex. It contains the event record described in "EventIndex Record Contents" like the run number, event number, trigger mask, time of data taking, etc.

The EventIndex file contains one Header message at the beginning and one Trailer message at the end. Between them, one or several sequences of processed files which begin with BeginGUID and end with EndGUID. For each processed file one or more TriggerMenu messages and a sequence of EIEvent records are written.

Although the protocol buffer format tries to store the information using the least space possible, the compression factor obtained is greater than 86%, as consecutive EIEvent messages usually contain very similar (and partially equal) information, so the compressor can reduce the space very significantly.

## Supervisor

The Data Collection Supervisor is in charge of tracking all data collection steps. It validates the produced data and informs the consumers about the presence of data to be transferred into the Hadoop cluster at CERN. It also allows following the indexing progress of datasets and containers thanks to its web interface.

The Supervisor receives messages sent by each producer job with information about what has been processed. This information includes among other things: the dataset name, the task and job identifiers, the location of the produced object store, the GUIDs of the processed files, and the number of events and the number of different event identifiers processed per file. The supervisor collects this information, and it is thus able to know which datasets are being indexed by which tasks in which system.

As part of the file and dataset metadata information, Rucio stores the number of events that they contain. This information is used to track the progress of dataset indexation. Furthermore, since each job sends the number of events that it has processed per file, this information can be compared against the one provided by Rucio to identify possible inconsistencies.

The supervisor also retrieves the information of the indexing tasks before they achieve a final state. To do this, the supervisor has to contact and decode the information provided by two different monitoring systems: conTZole [32] if the tasks is running in Tier-0, and PanDA Monitoring [33] if the task is running through PanDA on the WLCG Grid. Both monitoring systems provide information on the progress of tasks, like the number of jobs, number of events processed, status of the task and jobs, etc. Once the task has reached a final successful status, all the collected information from the jobs, from the task monitoring system, and from Rucio can be cross-checked:

- Each successful job should have produced and sent a message to the supervisor.
- Each file should have been processed by at least a job.
- The number of processed and produced events per file should match the number of events in the file according to Rucio.

When those checks are satisfied, the supervisor can assure the completeness and correctness of the produced information; then a validation object is created and stored in the Object Store. Among other information, the validation object contains the URIs of the produced Object Store objects that allowed the validation, as well as, for each object, the list of the files that were processed in that job and should be considered as valid. The consumers are informed through the messaging system about the validation objects that they should consume; they first retrieve the validation objects, process them retrieving from the Object Store the information that should be consumed and put into Hadoop. When all data have been consumed, they notify the supervisor about it, signalling once again how many events have been consumed. With this last message, the supervisor can mark the dataset as indexed.

Inconsistencies in the number of events, unprocessed files, lost or delayed messages, can be identified thanks to this validation procedure. A dataset that is not validated is kept in a validation queue, where the validation will be retried in after receiving possible delayed messages.

With all these pieces of information from different systems the supervisor is able to:

- Monitor the progress of the tasks that index each dataset;
- Identify failed production tasks;
- Declare obsolete indexing tasks that have problems and are going to be replaced by other tasks;
- Detect if any messages were lost;
- Identify inconsistencies between the processed files and the information retrieved from Rucio;
- Complete missing pieces of information in Rucio in the rare cases when they occur;
- Notify, through email if needed, about datasets that have duplicated event numbers detected at job level, i.e. within the few files of the dataset that were processed by the same job;
- Identify failed data transfers due to the death or disconnections of the consumers from the brokers;

Figure 2 shows the data collection process with the interactions between the different components and the information flow.

## Consumers

The Consumers are in charge of storing the EventIndex data in the final data store. They run centrally at CERN and in the current system there are as many consumers as messaging brokers, as this is sufficient for the current production rates. They are stateless independent entities that can be scaled up in case of necessity.

Consumers wait for validation messages from the Supervisor, containing references to the actual EventIndex data objects to be ingested. These objects are read from the Object Store with data encoded with Protocol Buffers [30] format. The data are then formatted for the current production schema, and stored in Hadoop files. These files are organized in directories named after each dataset container, and the current granularity is to write a file per dataset, but this is also configurable in the validation object. Each file contains data organized by key containing RunNumber–EventNumber, and its related value encoded in a CSV schema with all the event information. Information about the status of the processing is communicated back to the Supervisor, starting with the acknowledgment of the request. When all the objects are consumed and the file is written, the result is sent back with a control message again. In case of any failure, details are included. It must be noted that the granularity of the validation data can vary from a single object reference, to thousand of them belonging to a particular dataset.

This procedure is repeated for all validation messages produced for a dataset container. At this point the validation of the complete dataset is possible, and a different control message will trigger the final validation of a dataset container. This procedure writes a control text file in the Hadoop file system, containing all the URLs of the individual dataset files that were validated. This file is used by Oracle ("Data Storage in Oracle") to know which data have to be imported.

An individual Consumer typically processes a mean of 15 kHz (events processed per second), and we have observed a maximum of 28 kHz. The current implementation of the Consumer is a multi-threaded Java program, with thread pools using the Future pattern [34] in order to save resources, and exploit parallelism among internal data access and transformation tasks. The setup of the Hadoop writing channels and the input/output largely dominates the processing time, with the CPU used on data mangling and schema transformation taking a small percentage of the time. The Consumer design allows to easily include new data sink plugins, and it has been extended to support new data backends like Kudu [35], and now HBase/Phoenix [12, 36].

## Data Storage in Hadoop and HBase

### Data Formats in Hadoop

All data are stored in Hadoop MapFile format [37] on the Hadoop file system provided by CERN. The MapFile format is a basic Hadoop storage format with two related SequenceFiles [38] (another basic Hadoop storage format), one with data, the other being an index. Both SequenceFiles consist of key:value pairs ordered by key. In the data file the values are the payload, in the index file the values are the positions of the keys in the data file. The index file contains a fraction of the keys so that it can be kept in memory. The MapFiles allow fast random data access by the key that we use to query the data.

Some MapFiles contain full Event Index records, others contain various derived entities and records. This mechanism is transparent to the users, as all MapFiles are treated in the same way. Search results are also stored as MapFiles to be available for later reuse.

All MapFiles are registered in the Catalog, which is implemented as an HBase [12] table. The Catalog contains all information about each MapFile, its status, properties, history and relations to other MapFiles.

MapFiles can be searched in three ways:

1. Key-based search on (sequence of intervals of) keys. This method gives almost immediate results. The primary key for MapFiles containing event records consists of the RunNumber–EventNumber pair, that is unique within each MapFile, as it corresponds to a dataset. Derived MapFiles can have different primary keys. Further, more detailed selections can follow after the key-based search.
2. Full Map/Reduce search. The search clause may contain Java code or complete Java classes implementing the Mapper and Reducer steps of the Map/Reduce process.
3. Full scan search. It is the slowest way, but it is useful to understand the details of the search process.

Most search and formatting options can contain any legal Java code using MapFile variables.

### Compression of MapFiles

Due to growing storage space, we had to consider compressing the data file of each MapFile.

In the record-compressed SequenceFile format, each record is compressed separately, but the keys are not compressed. With the default codec (the "deflate" format) the space savings are in our case a factor of 2 to 4, depending on the data type.

Using the block-compressed SequenceFile format, groups or blocks of keys and records are compressed together. The block size for compression - the size of uncompressed keys plus values that become compressed together - is configurable. In our case, with the default block size (128 MB) and compression codec we reduced the file size by a factor 10.

For any tool reading the files as standard MapFiles, the change of compression type of the SequenceFiles is transparent. In block-compressed format, to read one record one has to read the whole block; however, querying the data we did not see measurable differences relative to the record-compressed data files. We decided to use the block-compressed format, so that the average size per event was reduced to 100 bytes, with the largest datasets so far (100 million events) residing in 10 GB files, still a manageable size.

### Data Import to MapFiles and Copy to HBase

The Consumers ("Consumers") write in Hadoop one sequential file for each dataset. These files are converted into MapFile format and copied into their dedicated space. All those MapFiles are then registered in the Catalog, which is implemented as an HBase table.

After each successful dataset import, the event information (without trigger record, for space and performance reasons) is uploaded to an auxiliary HBase table for fast event lookup operations.

All import and search operations are also registered in the system journal, together with all relevant information. The journal is also implemented as an HBase table. The full data flow within the Hadoop system is shown in Fig. 3.

### Duplicate Event Detection

After each import, the MapFiles consistency is verified and all potential problems are notified to the relevant users. The most important inconsistency is the presence of the events stored several times in the same dataset MapFile, which is usually a consequence of a problem in the previous stages of data processing. When duplicated events are found, the production managers are automatically notified by email.

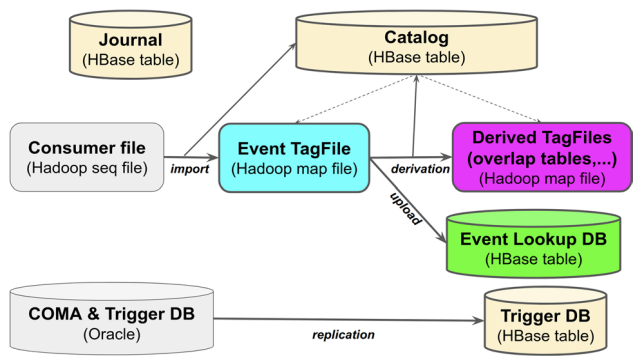

**Fig. 3** The overall data flow within the Hadoop system. The sequential files, one per dataset, written by the Consumers are imported into MapFile format and registered in the Catalog. A subset of the information is stored also in the event lookup table in HBase. Consistency checks are applied and derived information is saved also in Mapfiles. Trigger decoding information is imported from Oracle to HBase for local use. All actions are recorded in the journal

## Trigger Decoding

The trigger record for each event is transferred as a bitmap, where each bit corresponds to a trigger chain. In order to store the trigger data in Hadoop in an easily searchable and retrievable way, it has to be decoded with the help of the trigger mask for the given dataset, which in turn can be retrieved from the trigger database using the trigger key (SMK) of the dataset. The trigger tables are available in different databases: the COMA (COnditions MetadatA) database [39] contains all trigger information for real data and the MonteCarlo Trigger DB (TriggerDBMC) in Oracle contains the data for MC simulation.

The EventIndex replicates the trigger tables from COMA and TriggerDBMC to HBase tables and then uses them for trigger decoding [40]. If the SMK is absent in the event record, it is possible to obtain it from the run number for the real data and from the reconstruction tag for MonteCarlo simulation. This information is also replicated to the HBase tables in the Hadoop store.

The trigger decoding data flow is presented in Fig. 4. The trigger masks from event records are decoded [40] using HBase tables, converting chain counters to chain names. The list of trigger chain names obtained after decoding are then stored in updated event records. The information obtained is used for trigger-based selections or to calculate trigger overlaps, helping the trigger chain optimization.

## Derived Statistics and Correlation Tables

After the dataset import and successful verification, several derived tables are created automatically:

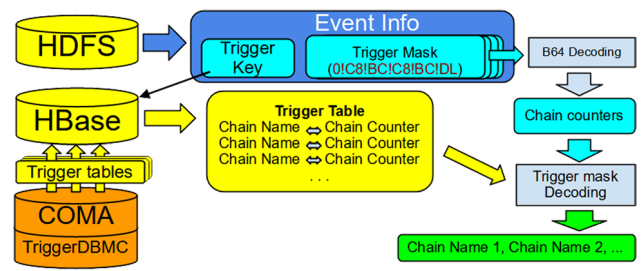

**Fig. 4** Trigger information decoding data flow. The EventIndex information in the Hadoop file system (HDFS) contains the trigger key for each event, which is used to retrieve the copy of the relevant trigger table stored in HBase. The event trigger mask is then decoded using this trigger table and the result is stored back with the event record in Hadoop

- The dataset overlap table contains the numbers of common events between different datasets in the dataset derivation chain.
- The trigger overlap table contains for each dataset the number of trigger chain pairs which were fired simultaneously.
- The trigger statistics table contains for each dataset the number of fired trigger chains of each type. While this table is created separately, it can be seen as a subtable of the trigger overlaps table.

All derived tables can be interrogated with the standard tools because they are implemented as normal MapFiles. Overlap tables can be also visualized as Graphs (see Fig. 5).

## Command Line Interface

Several commands were implemented to give access to the stored data:

- Catalog (catalog) to search and modify Catalog entries.
- EventIndex (ei) to search all datasets using either direct searches or complex Map/Reduce jobs. The EventIndex command allows the use of any legal Java code as a search or result clause.
- EventLookup (el) for fast search of the physical datasets corresponding to an event (specified as a pair of run number and event number).
- TriggerInfo (ti) to perform search and analyses of the trigger information.
- Inspector (inspect) to see the actual content of a MapFile.

Commands to access the EventIndex store are available directly in the CERN Hadoop cluster for the data

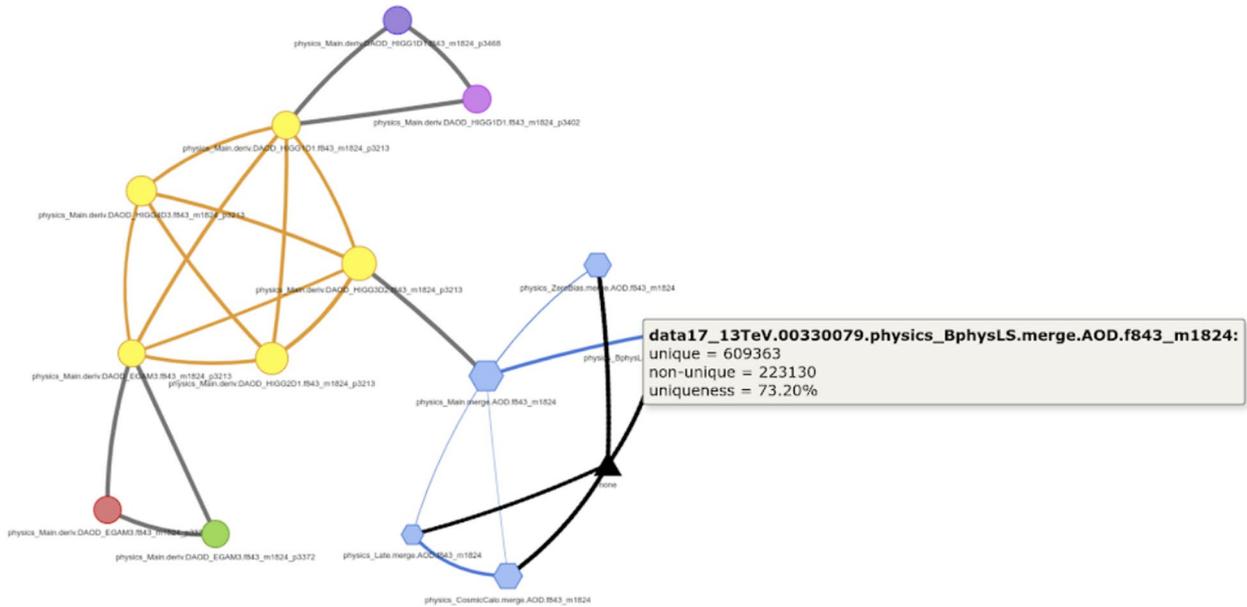

**Fig. 5** Screenshot of the EventIndex Graphical Web Service. It allows the navigation of graphs of data entities. Datasets with event overlaps are shown in this example. The graph can be further explored, other relations can be shown and operations on the vertices and edges can be executed—either from the web service itself or by calling other ATLAS services. The example shows possible actions available for a dataset and the tabular view of the dataset trigger statistics

management tasks. The (client) commands that do not modify the store contents are available to all ATLAS members in the standard locations, on the CERN Linux machines as well as in the ATLAS CVMFS [41] environment that is available world-wide. The remote invocations go through a Tomcat [42] based web service.

### Web Interface

All provided commands are also available via a Tomcat-based standard REST [43] Web Service. On top of the simple form-based interface corresponding to the command arguments, high level graphical and interactive Web Services have been implemented. They offer several graphical ways (histograms, Venn diagrams, Graphs, etc.) to display the data contents and their relations, as shown in Fig. 5.

### Event Lookup or "Event Picking"

Event lookup is the most important and heavily used functionality. It returns the GUIDs and optionally the stream type, dataset name and other parameters for user specified sets of real or simulated events, identified by run and event numbers. The search can be narrowed by specifying the trigger stream, data format and version. By default, the lookup is performed in the HBase table, the best performing back-end. It is also possible to run identical queries against the MapFiles from which the data are ingested into the HBase table, as described in "Data import to MapFiles and copy to HBase", and which were historically the first event lookup implementation.

### Comment on Free and Open-Source Software

When it became necessary to convert all MapFiles to the block compressed SequenceFile format (see "Compression of MapFiles"), it turned out that random access queries were not working on the block compressed files. Thanks to the availability of the source code, we were able to fully investigate the issue and track it down to a bug in a MapFile method. Having resolved the problem and incorporated the patched version into the EventIndex, we contributed the patch to the Hadoop project; the patch was accepted and we also had a chance to learn the Hadoop project practices.

The Hadoop project software is released under the Apache License 2.0 [44], which is a free software license according to the Free Software Definition [45], and this turned out to be a crucial point. Hadoop has created an efficient infrastructure and stimulating atmosphere for project contributors with easy access to the full build and test environment, the use of modern compilers and build tools, extensive use of unit testing and of advanced code quality assurance tools, systematic and consistent use of an issue tracker, an automated contribution testing system, and expert and friendly contribution reviewers.

It is worth mentioning that, apart from Oracle (see "Data Storage in Oracle"), all the software we use is free and open-source software.

## Data Storage in Oracle

The initial implementation of the EventIndex store in Hadoop showed several important shortcomings by the end of 2015, the first year of LHC Run 2. With the versions of Hadoop and HBase and the hardware setup provided by CERN that we could use at that time, all queries became substantially slower as the amount of stored data increased. Simple event lookup queries for 10 events started taking over 1 min instead of the expected sub-second response, and counting events across large datasets (100 million events) took tens of minutes.

In addition to optimizing the Hadoop cluster setup, it was then decided to explore the possibility of storing a subset of the real data information in an Oracle database, exploiting this well-known technology to support the most important and time-consuming use cases, primarily event picking for real data events. For each real data event, the event record without trigger information, which constitutes 80% of the data volume, is copied to an Oracle database. Locating this data in existing Oracle servers also allows us to easily connect to other metadata in existing complementary repositories like the COMA [39] and AMI [16] systems, which store metadata related to runs and datasets, respectively.

## Data Structures

A relational model was found to be well suited to the task [46]. The simple relationship of datasets to events lends itself to a very simple relational table structure as shown in Fig. 6. The two leftmost tables (in blue) store each indexed dataset and its events, respectively, driving the core functionality of the system as well as serving some secondary use cases. In ATLAS, datasets are uniquely identified by a string concatenating 6 fields, each of which is stored as a separate column in the *Datasets* table:

1. Project name: A string encoding the LHC beam type with the year of data taking,
2. RunID: The run number,
3. Stream name: Events passing specific triggers are written to one or more data streams,
4. Data format: The stage of processing at which the data are indexed (usually AOD),
5. AMI Tag: a string encoding the processing steps these events have undergone, i.e. effectively the processing version for the events in this run and stream.
6. Production Step: A short string to distinguish between an intermediate or final processing stage.

Since the Datasets table is the parent table for all other tables of the schema, this table has an integer primary key associated with each indexed dataset name as is common in relational database design.

For each dataset, all events are stored in the *Events* table. We store up to 3 GUIDs per event, which we found to be sufficient. References are numbered starting at 0 (the GUID of the indexed file), with subsequent references 1 and/or 2 used for its upstream file formats. Since available GUID types (RAW, AOD, and DAOD) are common to all events in the dataset, GUID types are stored only once in the Datasets table (another advantage of relational design).

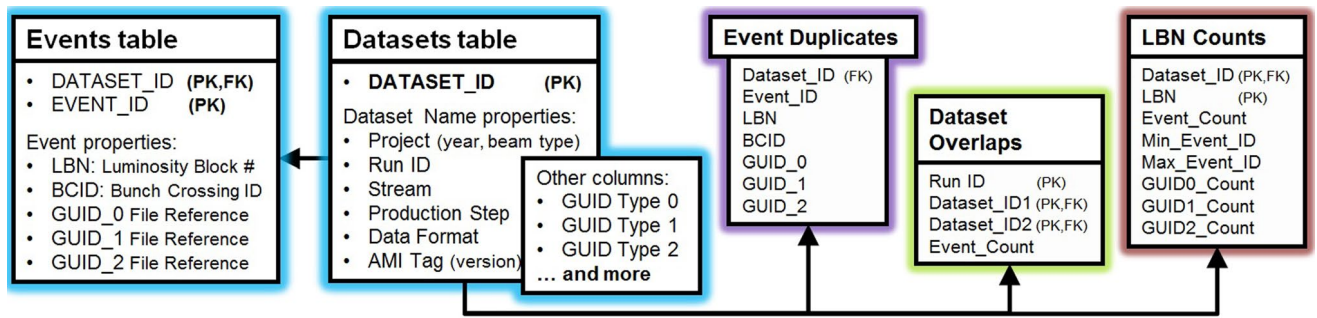

**Fig. 6** Relational tables of the EventIndex Oracle (EIO) schema. The "Datasets" table contains one row for each imported dataset. The unique events for each dataset are stored in the "Events" tables and its duplicated events (if any) are stored in the "Event Duplicates" table. Aggregated information about each dataset is stored in "LBN Counts" (a count of events per Luminosity Block) and "Dataset Overlaps" (the number of common events between datasets of the same run number)

Also, at the event level, we store the event's luminosity block number (LBN) and LHC bunch crossing identifier (BCID), which are useful for use cases described in "Web Interface".

The *Event Duplicates* table keeps the list of all duplicated events found. Duplication can (and has been shown to) occur at any stage in processing. In the initial 2015 loading of the data, hundreds of datasets were found with duplicate events; using data stored in this table combined with event counts at each stage of processing (from the AMI database), we could identify the stages at which duplication occurred. Subsequent refinements in upstream systems have considerably reduced the occurrence of duplicated events in current data.

The *Dataset Overlaps* table stores the number of events in common between different datasets of the same run. This data serve a secondary use case of providing these overlaps to experts in DAOD production for refinement of the ATLAS Derivation Framework [47].

The *LBN Counts* table stores the event counts and the number of associated unique GUIDs by LBN. Data in this table have multiple secondary uses including many forms of integrity checks, determining the probable LBN for an event, investigating missing events and files, and understanding the splitting of luminosity blocks at file boundaries.

Both the 'Overlaps' and the 'LBN Counts' table content could be computed dynamically using the data in the two primary tables, but we chose to materialize this data in these tables since the computation can take more than a few seconds, the data volume is minimal, and some of the aggregated data are used in multiple services.

When a new dataset appears in Hadoop storage, it is considered for import into Oracle if the run exists in the COMA system (which contains only runs of potential physics interest), and if the Stream meets similar selection criteria (e.g. there are no known use cases for indexing datasets in calibration streams). If the dataset passes these requirements, the "Import process" stage imports the dataset and its events into "Staging tables" which are similar to the final tables, but without indexes or constraints. The "Oracle scheduler Jobs" stage runs verification checks such as checking that the events are consistent with belonging in the same dataset and flagging if any duplicate events are found. If the data pass verification checks, the data are moved to the "Destination Tables", writing any duplicate events to a separate table, while keeping one copy for the Events table.

Subsequently, supplemental information is added to the Datasets table including data from other repositories (COMA and AMI) as well as aggregated information from Events table loading. This includes dataset status flags, various relevant dates, event counts both within the system as well as related counts in ATLAS file systems (upstream dataset files), counts of unique GUIDs associated with the dataset, and counts of total and unique duplicated events. In addition, the datasets with Run, Stream, and Data Format in common are ranked by dataset creation date, which is useful to help users find the latest processing of a set of events. These columns are used to enhance various services, and/or are included in user interfaces and reports.

A number of additional database optimization techniques deployed in this system, which further minimize storage volume (beyond relational normalization mentioned previously), transaction volume and database load, and optimize query performance for use cases, strongly deserve mention:

– The Events table is "list" partitioned by DATASET_ID. The main advantage is that sets of events can be deleted by simply dropping the associated partition. This operation is needed more often than we initially expected because datasets are sometimes re-indexed because of constituent file loss on the grid (which invalidates the associated GUIDs).
– For the Events table we use Oracle's "basic" compression for table data and key compression on its primary key index. Moreover for data loading we use Oracle's direct data load interface. In combination, storage utilization is reduced by a factor of about 3.5 which has the added advantage of reducing similarly the I/O footprint for writing data, undo and redo to the storage subsystem.
– Up to three GUID reference columns per event in the source data are 36-character strings (for example "21EC2020-3AEA-4069-A2DD-08002B30309D"). In our Events table, we store these columns using the non-standard "RAW" data type, reducing the 36 bytes of storage per GUID to 16 bytes. This considerably decreases the Events table per-row volume without loss of functionality: when the GUID columns are queried, an Oracle function easily converts them back to the original CHAR type (event lookup is always by EventID, not by GUID).

After optimization, the storage volume is ~20 bytes per event, a factor of 10 reduction from the initial 210 bytes per event for this data imported from Hadoop. This reduction, however, is only for the table segments. Adding the primary key index overhead, the reduction factor drops to 5 (the size of the parent table is negligible, only 8 MB). So overall, including indexes, storing $25 \times 10^9$ events requires less than 1 TB of space (rather than 5 TB). The savings of 4 TB of disk space, in itself, is not the foremost point but has a knock-on effect which is particularly beneficial for query performance: it enables the caching of a larger fraction of the database rows into the database data cache (buffer pool) which yields real performance gains in query response (around 10 ms for simple queries). In summary, using a relational model and a number of carefully chosen techniques available in Oracle RDBMS results in an impressive minimization of resources while exceeding performance goals.

## Web Interface

The central part of the user interface is the EIO (Event Index in Oracle) Browser shown in Fig. 7, which allows users to easily find indexed datasets and their properties.

The browser offers dynamic filtering of datasets by any of the dataset name fields and/or other dataset characteristics. With each iteration of selection criteria, the system shows the number of remaining datasets meeting the criteria and displays the remaining criteria. Once the user has selected their dataset(s) of interest, they can choose from the following services:

- *Event Lookup* serves the primary use of returning GUIDs for user specified events (RunID/EventID pairs). In the absence of a user specified dataset version, GUIDs from the highest EIO-derived ranked dataset are returned. The report provides additional details about the events found as well as information to help determine why events were not found.
- The *Dataset Report* includes a table displaying details about each selected dataset: the collected and derived information in the Datasets table. EIO event counts are compared to counts of the corresponding upstream dataset files which helps to understand event losses/filtering at each stage of processing. Links are provided to related AMI dataset and COMA run reports and to other EIO services described herein.
- The *Dataset Overlaps Report* shows the count and percentage of events in common between selected datasets of any run that is useful for the resource optimization of the offline production of DAOD [47]. Results are displayed in a colour-enhanced 2-D matrix (as in Fig. 8) showing datasets which overlap by more than a 70% threshold. This threshold and a choice of two overlap computation algorithms are configurable in the interface.
- The *Duplicate Event Report* displays all copies of events with any duplicated event identifiers in a dataset. It shows clearly the LBN(s) where duplication occurred and the associated GUIDs, from which, combined with event counts at each stage in processing, we can unambiguously determine the processing stage in which the duplication occurred.
- A *Missing Event Report* can be generated when a dataset has fewer *unique* events than expected. The cause may be intentional filtering or an unintentional error resulting in in-file event loss or entire files of events being lost; reports show event counts (and computed losses) at each processing stage. If those events have been completely processed and indexed in another version of processing, the report shows lost event ranges and associated LBN(s).
- The *Count by BCID Report* displays event counts in each LHC bunch crossing (BCID). During collision operations, one clearly observes the correlation in the peaks of recorded events with BCID with the LHC fill configuration of the run.

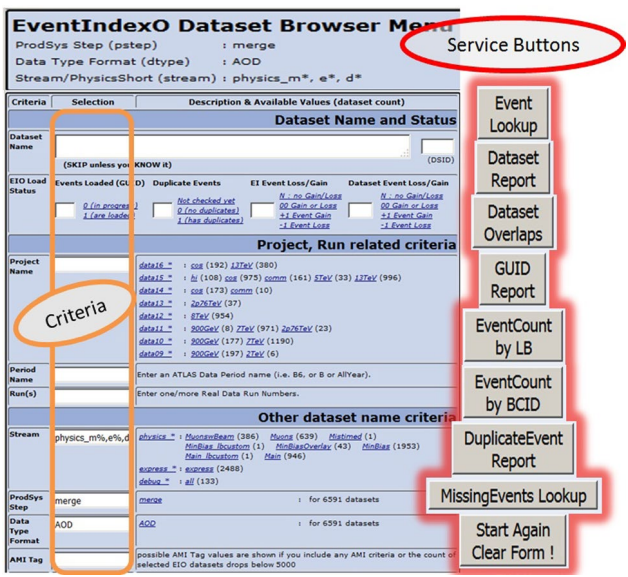

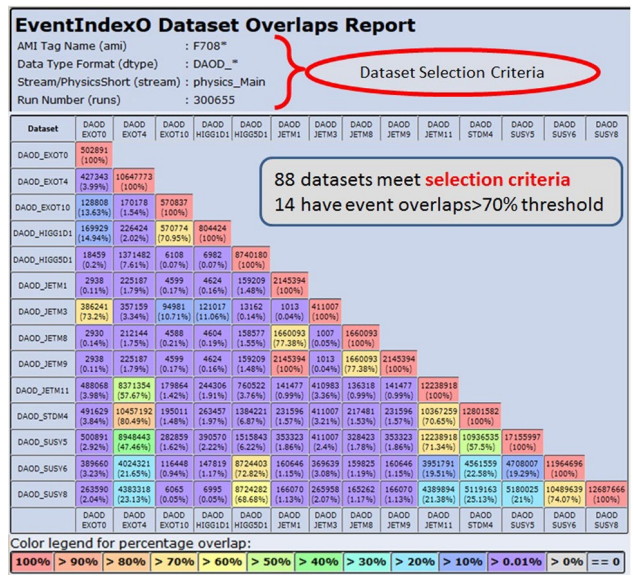

**Fig. 7** The EIO dataset browser entry page. Users can set search filters by clicking on predefined options or typing in the text boxes on the left, and then select the kind of report by choosing from the menu on the right. The reports are described in "Web Interface"

**Fig. 8** The EIO dataset overlaps report, as an example of the functionalities provided by the EIO dataset browser. This report shows the count and percentages of events in common between selected datasets, as described in "Web interface"

- The *Count by LB* and *GUID Reports* both display event and GUID counts along with EventID ranges per LBN which have been aggregated in the LBN Counts table. The GUID Report further shows the distribution of LBNs by GUID for selected LBN ranges: this is a useful cross check of event in-file metadata, which on occasion had incorrect counts causing problems in the ATLAS luminosity accounting software. This report has been helpful to identify incorrect in-file metadata since this system gets this information via a completely different path.

## System Monitoring

The successful operation of the EventIndex system depends on a number of different components. Each component has different sets of parameters and states and requires a dedicated approach for monitoring. A first version of the EventIndex monitoring tools [22] based on Kibana [48] was developed in late 2014, but it suffered from performance issues, so a new version based on InfluxDB [49] as data store and Grafana [50] for the display was developed [51].

The monitoring infrastructure consists of two parts, producer and viewer. The producer part is responsible for collecting data and transferring them to the database; it includes the scheduler, a number of Python scripts and the database. The scheduler uses a cron utility to run the Python scripts at fixed times. The Python scripts collect data from CERN and Grid sites and insert them into the database. Several different modules monitor different system sub-components: Data Production, Consumer Processes, Hadoop Imports, Hadoop Cluster, Trigger Database, Web Interface, Server Status, Event Picking Tests and Data Volumes. Each of these modules requires a different approach for data collection and processing, thus every module has its own Python script and scheduler to run it. The viewer part is responsible for the graphical presentation of data. Figure 9 shows the functional schema of the EventIndex monitoring system and the data flow.

Grafana supports different back-end databases; it was decided to use InfluxDB as a front-end database because support for InfluxDB+Grafana is provided by the CERN-IT Monitoring group. Although the group policy does not allow writing data directly to InfluxDB, an HTTP endpoint to the middleware that moves data to the database is provided for records in JSON format. This JSON format has a common part that is the same for all databases supported by the CERN-IT Monitoring group, and custom parts that are different for each database and carry the specific information for each service to be monitored.

The visualization component has a status dashboard for all modules, dashboards for the most important parameters of each module and links to the module details pages. The current status for each module is calculated using its own algorithm based on the module critical parameters. The status can have one of following values:

- "available" (green)—the module works correctly
- "degraded" (yellow)—the module has some non-critical problem
- "unavailable" (red)—the module has a critical problem
- "N/A" (white)—monitoring data are not available for this module.

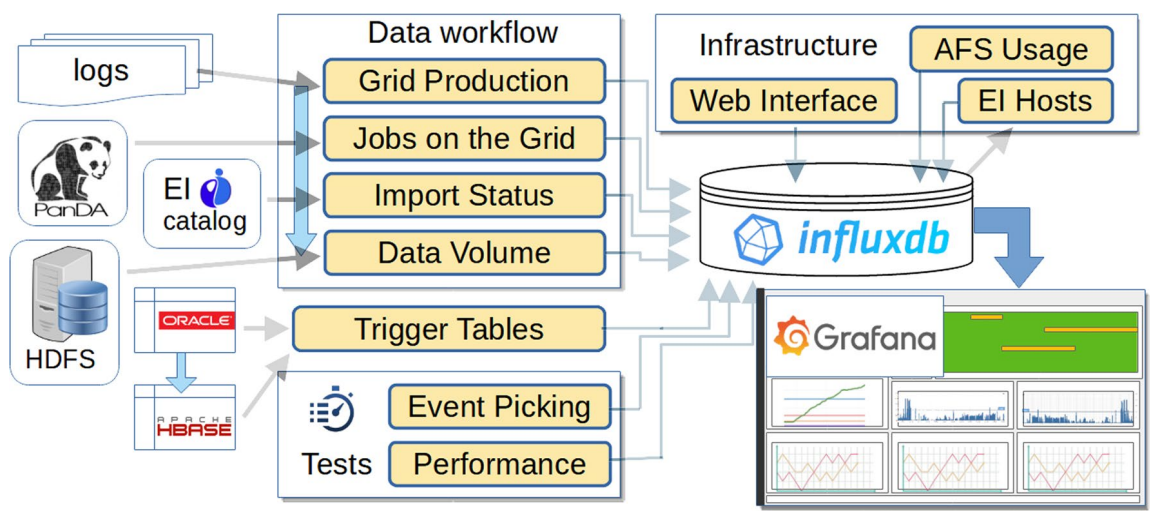

**Fig. 9** Functional schema for the monitoring system of the EventIndex components. Several modules collect information from multiple sources about the status of EventIndex processes (top-left), functional and performance tests (bottom-left) and the computing infrastructure (top-right) and store this information in an InfluxDB database; the data are then displayed in Grafana dashboards

The details page of each component usually contains additional dashboards. The status of each EventIndex service is also fed into the global ATLAS service monitoring view that the computing operation shifters check periodically; in case of problems, the experts are notified and can intervene promptly.

## Operations and Performance

Before the start of LHC Run 2 operations, we indexed all LHC Run 1 datasets in AOD format on the Tier-0 cluster. In this way we collected information on the RAW dataset provenance and on the trigger parameters of each Run 1 event. After that, real time operations started. The indexing jobs are distributed on all sites available to ATLAS, selecting primarily those that are closest to the input dataset location from the network point of view, as shown in Fig. 10. The indexing jobs are fast, as the producer transformation only reads the header of each event and takes between 10 and 50 ms/event, depending on whether the trigger record is needed or not; each job indexes several files and runs for 30 to 60 min. The total CPU consumption of EventIndex Producer jobs is well below $10^{-4}$ of the total CPU power used by the ATLAS experiment world-wide.

The percentage of jobs failing has been around 7%, with the main causes of errors being problems related to the input files and the sites operation, such as corrupted files, sites with storage or disk configuration issues, stage-in problems, etc. The indexing jobs are the first jobs run on just-produced datasets, so they are useful to detect at an early stage any problem with data corruption or unavailability. After contacting site administrators and the ATLAS data management operations team, the problems are usually solved promptly, so that simply re-running the problematic jobs is sufficient to achieve consistency.

The number of stored event records increased approximately linearly as shown in Fig. 11. Some datasets, especially those with type DAOD, do not have an infinite lifetime but are periodically replaced by newer versions generated with better calibrations or improved algorithmic code; after some time the old versions are deleted. The down steps in Fig. 11 correspond to periodic clean-up operations that remove the information regarding obsoleted datasets.

The current amount of disk space used by the Event Index data in the Hadoop cluster is shown in Fig. 12. Most of the disk space is used by real data and MC AODs, which contain the trigger records. The event generator datasets (EVNT) and derived analysis formats (DAOD) contain many more event records, but without trigger information, as it is either not yet available (in case of the generator-level EVNT datasets) or retrievable from the corresponding AOD datasets (for the derived formats DAOD).

The Hadoop system runs a variety of tasks, importing and cataloguing data, running consistency checks, establishing links between related datasets, and last but not least responding to user queries. Figure 13 shows the daily access statistics of the major Hadoop services. Accesses count all data handling procedures, including data import, user queries and functional tests.

Figure 14 shows the response times of the Hadoop server to event lookup queries selecting 10, 100, 1000, 10k and 50k events out of a dataset with one million records as a function of time. The event lookup is performed through the el client command, selecting randomly different events each time in order to avoid using cached results. The occasional glitches are due to other activities on the servers at the time of the queries. The response times are dominated by the query time for low numbers of events, and by the transmission time of the output for large numbers of events.

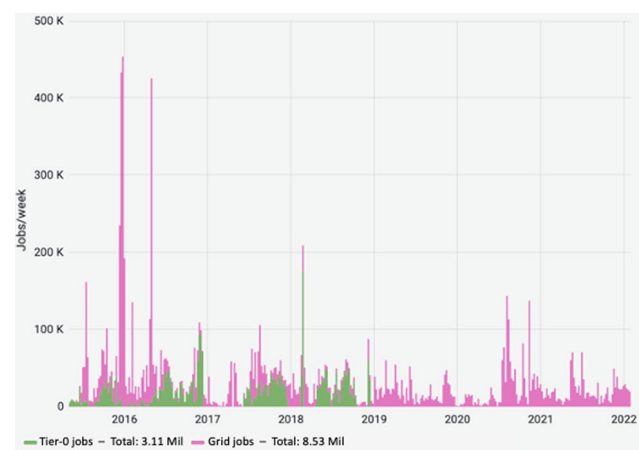

**Fig. 10** Distribution of the EventIndex Producer jobs run each week between the start of operations in May 2015 and January 2022. Jobs run on the CERN Tier-0 system that indexes all real data as soon as they are produced and reconstructed are indicated in green; jobs run world-wide on the ATLAS Grid resources are shown in purple

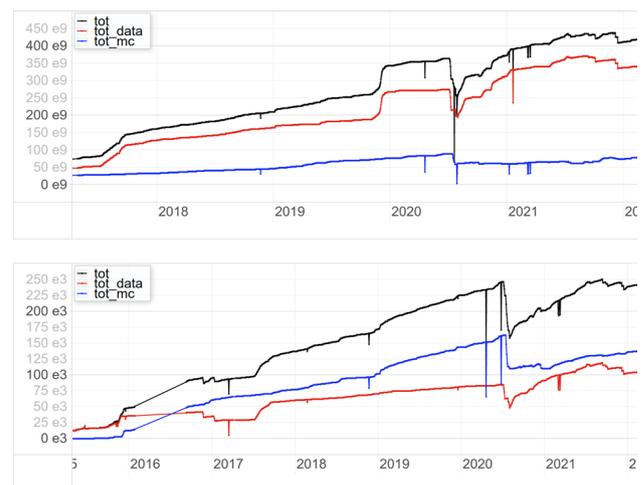

**Fig. 11** Event records (top) and datasets (bottom) stored in the Hadoop system between May 2015 and February 2022. Each plot shows separately real data in red, simulated data in blue and their sum in black

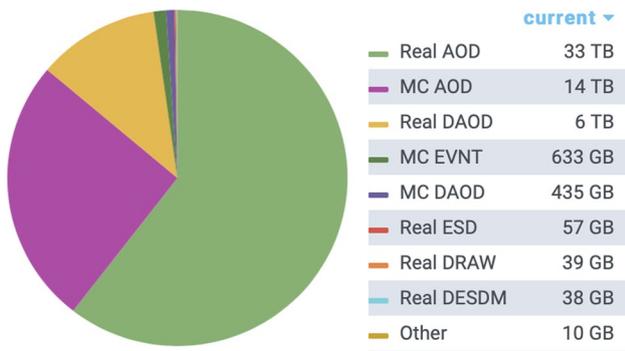

**Fig. 12** Data volume used in the Hadoop cluster, split by data type, June 2021

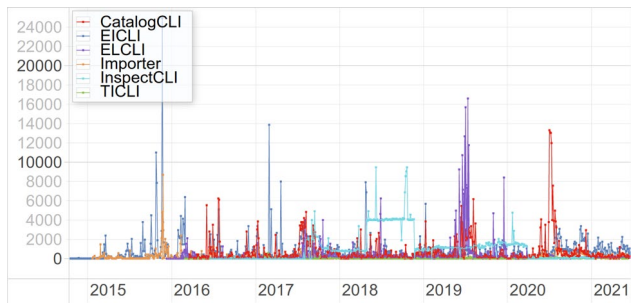

**Fig. 13** Access statistics of the Hadoop system between May 2015 and June 2021. The statistics is dominated by internal processes, like data imports, event counts and consistency checks, plus the regular functional and performance tests

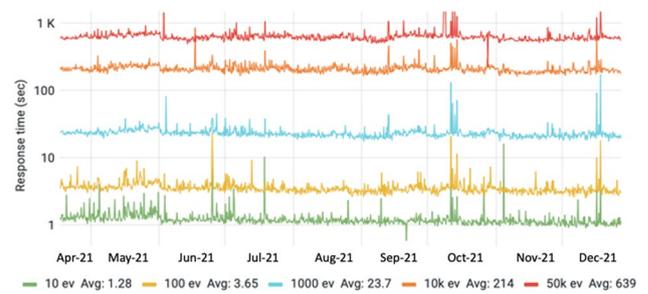

**Fig. 14** Response times of the EventIndex Hadoop server to event lookup queries selecting 10, 100, 1000, 10k and 50k events out of a dataset with 1 million records as a function of time, recorded between April and December 2021. Note the occasional longer response times due to other activities in the Hadoop cluster

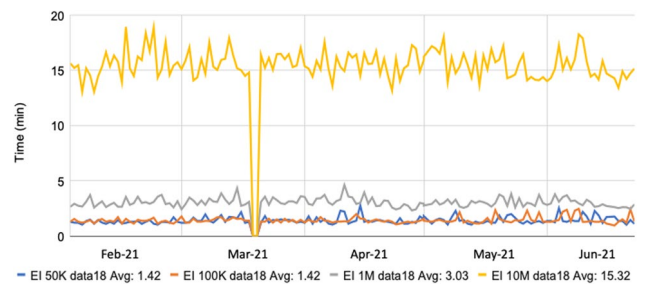

**Fig. 15** Response times of the EventIndex Hadoop server to queries using Map/Reduce jobs to retrieve information on all events from datasets containing 50k, 100k, 1M and 10M events as a function of time, recorded between February and June 2021. The discontinuity corresponds to a day of Hadoop cluster maintenance, during which no statistics were collected

The response times of the Hadoop server to queries retrieving information on all events from datasets containing 50k, 100k, 1M and 10M events are shown in Fig. 15 as a function of time. These queries are performed through the `ei` client command. The response times are dominated by the setup time of the Map/Reduce job for low numbers of events, and by the transmission time of the output record for large numbers of events.

The EventIndex data stored in Oracle increase in size in parallel to the growth of the main Hadoop store, as shown in Fig. 16. Due to the relational database nature of Oracle, a large amount of indexes is stored together with the actual payload data, so that a similar amount of disk space is used by actual payload data and the indexes.

Most of the event picking requests are for single events in RAW format; other requests are placed from time to time from physics analysis groups who need to extract their complete highly selected data sample for further processing and/or more detailed analyses; so far the EventIndex system could cope very well with all requests.

Figure 17 shows the statistics of the event picking jobs run between January 2019 and June 2021. During this period 9.5% of the jobs were automatically "closed" and rescheduled to another site while waiting for the input file (or files) to be staged from tape, and 6.5% of the jobs failed after waiting for the input file(s) for more than 3 days; resubmitting the same jobs normally works, as the wait time is then doubled. The tape reading queues work in FIFO mode, so it is not possible to assign a higher priority to tasks requesting a single file as opposed to staging large datasets needed by production activities.

## System Evolution

The described storage implementation reflects the state of the art for BigData storage tools in 2012–2013 when the project started, but several different options appeared since, even within the Hadoop ecosystem. With the increase of data-taking and simulation production rates foreseen for LHC Run 3 (2022–2025) and even more for LHC Run 4 (High-Luminosity LHC, from 2028 onwards), a re-design of the core systems is needed. In order to be safe, a new system should be able to absorb a factor 10 more event rate than the current one, i.e. 100 billion real events and 300 billion simulated events each year.

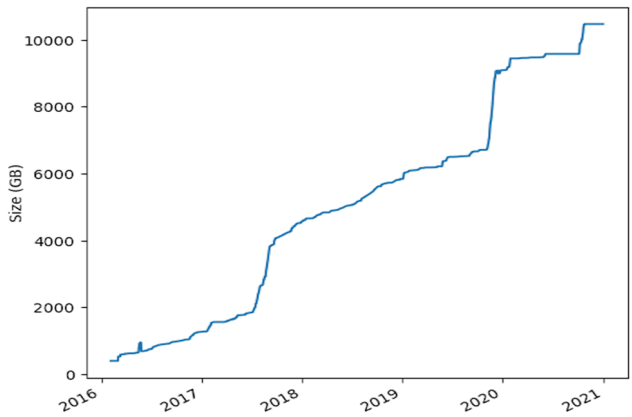

**Fig. 16** Disk storage size used by EventIndex tables in the Oracle cluster. Almost half of the disk space is taken by the large amount of indexes stored alongside the main data to optimize the read performance

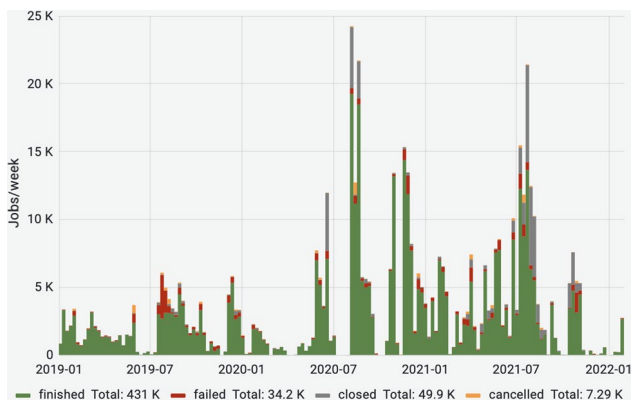

**Fig. 17** Event picking jobs run each week world-wide between January 2019 and January 2022

Investigations on several structured storage formats for the main EventIndex data to replace the Hadoop MapFiles started a few years ago [52]. Initially it looked like Apache Kudu [35] would be a good solution, as it joins BigData storage performance with SQL query capabilities [53]. Unfortunately Kudu did not get a sufficiently large support in the open-source community and CERN decided not to invest hardware and human resources in this technology.

HBase had been evaluated as the main data store at the beginning of the project, but was discarded at that time because of performance restrictions. Nowadays instead, it is able to hold the large amounts of data to be recorded, with a much-improved data ingestion and query performance thanks to the increased parallelization of all operations. Additional tools like Apache Phoenix [36] can provide SQL access to HBase tables, if the tables are designed appropriately upfront, which can be done in our case.

HBase works best for random access, which is perfect for the event picking use case where we want low latency access to a particular event to get its location information. Use cases where we need information retrieval (trigger info, provenance) for particular events are served by fast HBase gets, with good performance. In addition, analytic use cases where we need to access a range of event information for one or several datasets (derivation or trigger overlaps calculation), can be solved with scans on these data. They can be optimized with a careful table and key design in order to maintain related data close within the storage, reducing access time.

HBase is a column-family grouped key:value store, so we can benefit from dividing the event information in different families according to the data accessed in separated use cases; for example we can maintain event location, provenance, and trigger information in different families. Further analytic use cases on larger amounts of data are not foreseen, but still can be achieved running Map/Reduce or Spark jobs on the HBase files, as they are stored in the Hadoop file system.

Apache Phoenix is a layer over HBase that enables SQL access and provides an easy entry point for users and other applications. Although HBase is a schema-less storage, Apache Phoenix requires a schema and data typing to provide its SQL functionalities; nevertheless schema versioning and dynamic late binding for the same tables are supported as well.

EventIndex data rarely need schema changes, so we can benefit from Phoenix access, designing the required schema and tables accordingly. The table schemas and their relations [54] closely resemble those implemented for the Oracle version of the data store ("Data Storage in Oracle").

While updating the core storage system, other components can be revised and if necessary updated or replaced:

– The Producer implementation is currently done in python with a single thread. It will be upgraded to work with the latest data analysis software and external libraries like stomp.py [55], boto [56] and Protocol Buffers [30].
– The Data Collection system will use modern data processing technologies like Spark [57]. It will also allow to simplify all procedures, reducing data duplication and using common job management tools over the stored data.
– The Supervisor will be expanded to cover the entire workflow, from the selection of datasets to be indexed to the storage of data in HBase.
– The detection of duplicated events and the calculation of statistics for each dataset will be done "on the fly" during the import process.
– A new implementation of the Trigger Counter will make direct use of the Hbase/Phoenix infrastructure, which provides fields and families to store the six trigger masks of the event.
– A graph database layer working on top of any SQL database has been implemented to deliver a graphical and highly interactive view of the EventIndex data stored in the Phoenix SQL database. Thanks to its SQL genericity, this layer can work with all ATLAS data stored in SQL databases, thus providing a global navigable overview of all ATLAS data. All data are accessed directly via the standard *Gremlin* API [58] and the interactive graphical Web Service.

A prototype of the new storage and associated systems showed timing performances for data ingestion and for lookup well within our specifications.

A new tool was developed since 2021: the Event Picking Service [14]. It consists in a web service that can receive a list of events to be retrieved, with some optional specifications like the trigger stream and the data type to search for, and it will take care of all operations that were previously done by hand: query the EventIndex store to find the GUIDs of the files containing these events, submit the PanDA jobs to retrieve the events, retry the jobs if necessary, store the outputs in a central and safe location, inform the requester of the status of operations. It is useful to submit "massive" event picking requests, with numbers of requested events in excess of 10 thousand, for particular physics analyses that require dedicated reconstruction processes to be run on relatively small samples of pre-selected events.

# Conclusions

The ATLAS EventIndex was designed to hold the catalogue of all ATLAS events in advance of LHC Run 2 in 2012–2013, and all system components were developed and deployed in their first implementation as described in this paper by the start of Run 2 in 2015. As any software project, it went through several stages of development and optimization through the years. Thanks to the partitioned project architecture, each new component version could be tested in parallel with the production version and phased in when its performance was considered stable, and better than the previous version. The EventIndex operation and performance during and after the LHC Run 2 period has been satisfactory.

The significant increases in the data rates expected in LHC Run 3 and the subsequent HL-LHC runs required a transition to a new technology for the main EventIndex data store. A new prototype based on HBase event tables and queries through Apache Phoenix was tested and showed encouraging results. A good table schema was designed and the basic functionality was ready for operation in advance of the start of LHC Run 3 in 2022. We are now working towards improved performance and better interfaces; according to our expectations, this system will be able to withstand the input data rates foreseen for LHC Run 4 and beyond.

**Acknowledgements** This work was done as part of the distributed computing and databases applications research and development programme of the ATLAS Collaboration, and we thank the collaboration for its support and cooperation. Part of this work was funded through PRIN Project "STOA-LHC 20108T4XTM", CUP: I11J12000080001, of the Italian Ministry of Education, University and Research (MIUR). In Spain, MICINN partially supported this work under grants FPA2016-75141-C2-1-R and PID2019-104301RB-C21, which include FEDER funds from the European Union.

**Author contributions** All authors contributed to the development of the EventIndex software and system operations. All authors reviewed the manuscript.

**Funding** Open access funding provided by CERN (European Organization for Nuclear Research)

## Declarations

**Conflict of interests** The authors declare that they have no competing interests.